\documentclass[12pt,oneside,english]{article}
\usepackage{amsthm, amsmath, amsfonts, amsxtra, amssymb, euscript, mathrsfs, MnSymbol, verbatim, enumerate, multicol, multirow, color,babel, geometry, tikz,tikz-cd, tikz-3dplot, tkz-graph, array, enumitem, hyperref, thm-restate, thmtools, datetime, graphicx, tensor, braket, slashed, adjustbox, mathtools, bbding, esint, pgfplots, ytableau, float, rotating, mathdots, savesym, wasysym, amscd, pifont, setspace, wrapfig, picture, subcaption, caption, tabularx,youngtab,parskip}
\usepackage[numbers,sort&compress]{natbib}
\usepackage[normalem]{ulem} 
\usepackage[utf8]{inputenc}
\usepackage[all]{xy}
\numberwithin{equation}{section}
\numberwithin{figure}{section}
\usetikzlibrary{arrows, positioning, decorations.pathmorphing, decorations.markings, decorations.pathreplacing, decorations.markings, matrix, patterns, shapes}
\tikzset{
on each segment/.style={
decorate,
decoration={
show path construction,
moveto code={},
lineto code={
\path [#1]
(\tikzinputsegmentfirst) -- (\tikzinputsegmentlast);
},
curveto code={
\path [#1] (\tikzinputsegmentfirst)
.. controls
(\tikzinputsegmentsupporta) and (\tikzinputsegmentsupportb)
..
(\tikzinputsegmentlast);
},
closepath code={
\path [#1]
(\tikzinputsegmentfirst) -- (\tikzinputsegmentlast);
},
},
},
mid arrow/.style={postaction={decorate,decoration={markings,mark=at position .7 with {\arrow[#1]{stealth}}}}},
}

\hypersetup{colorlinks=true}
\hypersetup{linkcolor=black}
\hypersetup{citecolor=black}
\hypersetup{urlcolor=black}

\makeatletter
\def\oversortoftilde#1{\mathop{\vbox{\m@th\ialign{##\crcr\noalign{\kern3\p@}%
				\sortoftildefill\crcr\noalign{\kern3\p@\nointerlineskip}%
				$\hfil\displaystyle{#1}\hfil$\crcr}}}\limits}

\def\sortoftildefill{$\m@th \setbox\z@\hbox{$-$}%
	\braceld\leaders\vrule \@height\ht\z@ \@depth\z@\hfill\braceru$}

\makeatother

\theoremstyle{plain}
\newtheorem*{thm*}{Theorem}
\newtheorem{thm}{Theorem}[section]

\newtheorem{hypo}[thm]{Hypothesis}

\theoremstyle{definition}

\newtheorem*{defn*}{Definition}

\newtheorem*{thm:BU}{Theorem \ref{thm:BU}}
\newtheorem*{thm:BUpure}{Theorem \ref{thm:BUpure}}

\makeatother

\newcommand{\calh}{\mathcal{H}}

\begin{document}

\begin{titlepage}
\vspace*{-3cm} 
\begin{flushright}
{\tt CALT-TH-2020-029}\\
\end{flushright}
\begin{center}
\vspace{1.9cm}
{\LARGE\bfseries Holographic baby universes: \\ an observable story \\  }
\vspace{1.2cm}
{\large
Elliott Gesteau$^{1,2}$ and Monica Jinwoo Kang$^3$\\}
\vspace{.6cm}
{ $^1$ Département de Mathématiques et Applications, Ecole Normale Supérieure}\par\vspace{-.3cm}
{Paris, 75005, France}\par
{ $^2$ Perimeter Institute for Theoretical Physics}\par\vspace{-.3cm}
{Waterloo, Ontario N2L 2Y5, Canada}\par
\vspace{.2cm}
{ $^3$ Walter Burke Institute for Theoretical Physics, California Institute of Technology}\par\vspace{-.3cm}
{  Pasadena, CA 91125, U.S.A.}\par
\vspace{.2cm}
\vspace{.6cm}

\scalebox{.95}{\tt  egesteau@perimeterinstitute.ca, monica@caltech.edu}\par
\vspace{1.3cm}
{\bf{Abstract}}\\
\end{center}
{We formulate the baby universe construction rigorously by giving a primordial role to the algebra of observables of quantum gravity rather than the Hilbert space. Utilizing diffeomorphism invariance, we study baby universe creation and annihilation via change in topology. We then construct the algebra of boundary observables for holographic theories and show that it enhances to contain an `extra' Abelian tensor factor to describe the bulk in the quantum regime; via the gravitational path integral we realize this extra tensor factor, at the level of the Hilbert space, in the context of the GNS representation. We reformulate the necessary assumptions for the ``baby universe hypothesis" using the GNS representation. When the baby universe hypothesis is satisfied, we demonstrate that the ``miraculous cancellations" in the corresponding gravitational path integral have a natural explanation in terms of the character theory of Abelian $C^\ast$-algebras. We find the necessary and sufficient mathematical condition for the baby universe hypothesis to hold, and transcribe it into sufficient physical conditions. We find that they are incompatible with a baby universe formation that is influenced by any bulk process from the AdS/CFT correspondence. We illustrate our construction by applying it to two settings, which leads to a re-interpretion of some topological models of gravity, and to draw an analogy with the topological vacua of gauge theory.
}

\vfill 
\end{titlepage}

\tableofcontents
\newpage
\section{Introduction}

Recent developments on the AdS/CFT correspondence have suggested that some theories of quantum gravity are dual to an ensemble of theories, rather than a unique theory. In particular, in the case of the Sachdev--Ye--Kitaev model, Euclidean replica wormhole calculations for Jackiw--Teitelboim gravity in a multi-boundary AdS spacetime \cite{Maldacena:2016hyu,Penington:2019kki,Almheiri:2019qdq} have shed lights on some recent proposals to solve the black hole information paradox such as the island conjecture \cite{Almheiri:2019hni,Penington:2019kki}. What replica wormholes tell us is that, at least in low dimensions, the saddles of the Euclidean gravitational path integral carry a lot of information about the bulk, and in particular, about the Page curve of black hole evaporation. This phenomenon may not carry over to higher dimensions.

The success of these replica wormhole calculations generated considerable interest in multi-boundary holographic correspondences. In particular, in \cite{Marolf:2020xie}, Marolf and Maxfield proposed a model for Euclidean gravity in a spacetime with a fluctuating number of asymptotically AdS boundaries, which revived baby universe proposals coming from string theory \cite{Giddings:1987cg,Giddings:1988wv,Dijkgraaf:2006ab,Aganagic:2006je}. The main point of \cite{Marolf:2020xie} is that allowing topology changes in the bulk, by adding or removing a boundary, should affect the Hilbert space of the theory. They propose a construction for a Hilbert space of baby universes out of the gravitational path integral, and suggest that the diffeomorphism invariance drastically reduces the dimensionality of this Hilbert space, because of the quotienting of null states. More recently, McNamara and Vafa formulated a stronger baby universe hypothesis: the Hilbert space of baby universes in a unitary theory of quantum gravity in $d>3$ spacetime dimensions has dimension one \cite{McNamara:2020uza}.

In this paper, we show that the Marolf--Maxfield construction and the baby universe hypothesis can be much more precisely understood if the focus is shifted from Hilbert spaces to the algebra of observables of quantum gravity. In particular, we will show that the Marolf--Maxfield construction is a particular case of a standard operator algebraic construction called the Gelfand--Naimark--Segal (GNS) representation.  

In the context of holography, the Marolf--Maxfield construction implicitly relies on the Marolf--Wall thought experiment \cite{Marolf:2012xe}. The Marolf--Wall paradox was formulated quite early in the study of entanglement in quantum gravity, which remains crucial for various analyses in AdS/CFT. The main point of the Marolf--Wall thought experiment is that given a certain number of boundary theories, there is no way to know if they should be seen as independent theories that are dual to independent bulks, or as pieces of the same theory, that are allowed to form bulk wormholes with each other. In our setup, this will mean that the boundary observables will not be enough to describe the dual bulk in a setup where the number of boundaries can fluctuate. 

In particular, we construct an algebra of baby universe operators which acts nontrivially on the bulk theory, but trivially on the boundary theory. Diffeomorphism invariance will teach us that this algebra is well-defined and commutative. In order to take this bulk action into an account, an extra tensor factor will need to be added to the algebras of boundary observables.

We then utilize the gravitational path integral, and the standard operator algebraic technique of the GNS representation, to reproduce the Marolf--Maxfield construction and build the Hilbert space of baby universes, or more generally, of any commutative algebra of observables. In particular, we show that the commutative property of the baby universe algebra will indeed collapse the Hilbert space a lot: all irreducible GNS representations will always have dimension one, thus satisfying the baby universe hypothesis. 

When the baby universe hypothesis is satisfied, we demonstrate that the ``miraculous cancellations"\footnote{This concept is introduced in \cite{Marolf:2012xe} and was adapted in \cite{McNamara:2020uza}.} that happen in the corresponding gravitational path integral have a very natural explanation in terms of the character theory of Abelian $C^\ast$-algebras. This allows us to derive that the moduli space of pure restrictions of gravitational path integrals to the baby universe algebra corresponds to its Gelfand space. 

We then give a precise description of these gravitational path integrals in two different cases: the case of a baby universe algebra of Hermitian operators and the case where the baby universe algebra is generated by a group $G$. The former case enables an explicit formulation of the simple topological model given by Marolf--Maxfield in our framework, while the latter shows that a precise parallel analysis can be drawn between our construction and the QCD $\theta$-vacua.

We find that the baby universe hypothesis is valid if and only if the theory of baby universes is not described by an ensemble, which we formulate precisely in Theorem \ref{thm:BUpure}.
\medskip
\begin{thm}
\label{thm:BUpure}
Let $\mathcal{A}_{QG}$ be the algebra of observables of quantum gravity, and suppose that $\mathcal{A}_{QG}=\mathcal{A}_{res}\otimes \mathcal{A}_{baby}$, where $\mathcal{A}_{baby}$ is a commutative $C^\ast$-algebra. Let $\omega$ be a state on $\mathcal{A}_{QG}$. The GNS representation of $\mathcal{A}_{baby}$ induced by $\omega$ has dimension 1 if and only if the restriction of $\omega$ to $\mathcal{A}_{baby}$ is pure.
\end{thm}

Based on our further analysis, we give three physical conditions that imply the baby universe hypothesis to be true when put together. More precisely, we prove the following, Theorem \ref{thm:BU}.
\medskip
\begin{thm}
\label{thm:BU}
Let $\mathcal{A}_{QG}$ be the algebra of observables of bulk quantum gravity, and $\omega$ be a gravitational path integral state on $\mathcal{A}_{QG}$. Assume

\begin{enumerate}
\item for $\mathcal{A}_{baby}$ an Abelian $C^\ast$-algebra, $\mathcal{A}_{QG}=\mathcal{A}_{res}\otimes \mathcal{A}_{baby}$,
\item $\omega$ is a pure state on $\mathcal{A}_{QG}$,
\item $\omega$ can be factorized as $\omega=\omega_{res}\otimes\omega_{baby}$.
\end{enumerate}
Then, the GNS representation of $\mathcal{A}_{baby}$ is one-dimensional.
\end{thm}
$\mathcal{A}_{res}$ is a reservoir algebra of boundary theories, which will be defined in a precise manner in later sections.

The three assumptions underlying the baby universe hypothesis are quite strong. Assumption 1 is also necessary for the Marolf--Maxfield construction to work, and amounts to the validity of an analog of the Marolf--Wall proposal. We will show that it may require an approximate theorem for entanglement wedge reconstruction near a black hole horizon. Assumption 2 means that the theory of quantum gravity should not come from an ensemble average. Assumption 3 is the strongest one, and requires the baby universe observables and the boundary observables to have no interaction. We will show that some explicit models \cite{Giddings:1987cg,Dijkgraaf:2006ab} make this assumption break down.

\subsection{Organization of the paper}
The rest of the paper is organized as follows. 

In Section \ref{sec:baby}, we explain the history of baby universes in quantum gravity and string theory, and give general thoughts on Euclidean baby universes, and their relevance to construct a dynamical Lorentzian picture through the Wheeler--deWitt equation and modular flow. 

We discuss in Section \ref{sec:difftotopo} the consequences of diffeomorphism invariance for topological operations that amount to baby universe creation or annihilation. 

In Section \ref{sec:bulkalgebra}, we first describe the Marolf--Wall thought experiment, and then use it to show that the algebra of boundary observables must be enhanced by an Abelian extra tensor factor to fully describe the bulk at the quantum level. 

We utilize in Section \ref{sec:hilbertBU} the gravitational path integral to realize the Marolf--Maxfield construction as a GNS representation of this extra tensor factor. 

We turn to the case of the baby universe hypothesis in Section \ref{sec:BUH}. We first show in Section \ref{sec:purity} that the necessary and sufficient mathematical condition for it to be valid is when the gravitational path integral reduces to a pure state on the subalgebra of baby universe observables. Furthermore, in Section \ref{sec:newBUH}, we refine the baby universe hypothesis and derive a theorem with three sufficient physical conditions for it to hold. 

In Section \ref{sec:alpha}, we give a mathematical explanation of the ``miraculous cancellations" implied by this hypothesis from character theory. We elaborate further from this description to show that the moduli space of gravitational path integrals that satisfy the baby universe hypothesis (or $\alpha$-states) can be described by the Gelfand space of the algebra of baby universe operators.

In Section \ref{sec:examples}, we illustrate our results in two contexts. In Section \ref{sec:topo}, we revisit a topological theory given as an example in \cite{Marolf:2020xie} through the lens of the GNS representation. In Section \ref{sec:group}, we draw a precise analogy between our framework and the $\theta$-vacua of QCD, by considering the case where the baby universe algebra is generated by an Abelian group.

We explore in detail the three assumptions of Theorem \ref{thm:BU} in Section \ref{sec:assumptions}. The first assumption in Section \ref{sec:assumption1} underpins the Marolf--Maxfield construction, and amounts to the validity Marolf--Wall thought experiment, which boils down to a hypothetical result on approximate entanglement wedge reconstruction. The second assumption, which is explained in Section \ref{sec:assumption2}, amounts to accepting that a satisfactory theory of quantum gravity should not be described by an ensemble. Finally, the third assumption, which is in our opinion the most likely one to break down, is discussed in Section \ref{sec:assumption3} and it amounts to imposing that there is no possible interactions between baby universe observables and boundary observables. 

In Section \ref{sec:discussion}, we summarize and discuss the implications of our results, which questions the plausibility of the baby universe hypothesis in a realistic physical setup. We also clarify the issue of whether or not allowing for a fluctuating number of boundaries requires a modification of the traditional AdS/CFT correspondence.

\section{Baby universes and holographic theories}
\label{sec:baby}

The idea that baby universes can form in quantum gravity traces back to the pioneering work of Coleman \cite{Coleman:1988tj}. A baby universe is, roughly speaking, a new piece of a spacetime manifold that becomes attached to it, either in a connected way (wormholes can branch out) or in a disconnected way (isolated islands can form). This is due to the works of Giddings and Strominger \cite{Giddings:1988wv,Giddings:1988cx,Giddings:1987cg} that link the baby universe to the physics of spacetime wormholes. From the point of view of quantum theory, it follows that if the creation and annihilation of such objects is allowed, it has to somehow contribute to the Hilbert space of the theory.

In particular, utilizing tools from string theory, Giddings--Strominger proposed a third-quantized model which allows for topology changes in a Lorentzian spacetime. By coupling an axion field to gravity, they showed that at the quantum level, some topology changes could appear, and that the presence of an axion field could explicitly cause a baby universe to branch out.  A particularly remarkable consequence of this model is its predictive power, as it shows that the effective potential at low energy has a stable minimum for a vanishing cosmological constant. The cosmological constant problem was further linked to the emergence of wormholes by Preskill in \cite{Preskill:1988na}, and to the Wheeler--DeWitt equation, which provides a description of the dynamics of a superposition of large universes interacting with a surrounding wormhole baby universe gas \cite{Hebecker:2018ofv}. When taking a large universe to be the worldsheet of a fundamental string, the baby universe state is represented by the dynamical target space of string theory.

Baby universes are then formulated in the context of string theory in \cite{Dijkgraaf:2006ab,Aganagic:2006je}, where a precise construction using branes is built. In particular, the construction in \cite{Dijkgraaf:2006ab} is very explicit in a Euclidean description: the wave function of the multicenter black holes gets mapped to the Hartle--Hawking wave function of baby universes.

\subsection{Baby universes in Euclidean gravity} \label{sec:BUeuclidean}

We now turn to explaining how the recent revival of baby universes in Euclidean gravity could lead to a better understanding of Lorentzian quantum gravity in AdS/CFT. 

General Relativity on a Euclidean background is given by a Wick-rotated version of AdS spacetime, which is performed in \cite{DeWitt,DeWitt:2007mi}. In the case of a preferred time direction, the effect of the Wick rotation will be to transform the Schrödinger equation into the heat equation \cite{Wick}. When this time direction is compactified, the compactification radius can be interpreted as the inverse temperature. In particular, this Wick rotation operation preserves the gauge principle of diffeomorphism invariance, which is crucial for our purposes. 
Theories of quantum gravity usually have their vacuum defined in terms of a functional integral over spacetime metrics. If one wants to formulate a fully satisfactory 4-dimensional theory of quantum gravity, this functional integral should be performed over Lorentzian metrics, and look like
\begin{align}
Z:=\int_{g_{\mu\nu}}e^{iS(g_{\mu\nu})}\ ,
\end{align}
where $S$ is the Einstein--Hilbert action.

This ``manifestly Lorentzian" formalism based on a path integral runs into a lot of technical difficulties, as the space of Lorentzian metrics on a 4-manifold is very large in general, and positive-definiteness issues arise due to the presence of null vectors. An alternative is to instead try to look for a Hamiltonian formulation of the theory, which would define 4d quantum gravity as the time evolution of 3d Euclidean geometries in a preferred time direction. 

This is the basic idea of the definition of the Hartle--Hawking state \cite{HartleHawking,HartleHawking2}, which is a quantum superposition of 3d geometries. Such a state should satisfy the Wheeler--deWitt equation \cite{DeWitt1} 
\begin{align}
H\ket{\Psi}=0,
\end{align}
where $\ket{\Psi}$ is the Hartle--Hawking state, and $H$ is a Hamiltonian derived from the constraint equations of General Relativity. These constraints are derived from the Arnowitt--Deser--Misner (ADM) formalism \cite{ADM}.

The ADM formalism defines canonical variables for General Relativity. In particular, for a globally hyperbolic spacetime, one can consider a foliation of spacetime of the form 
\begin{align}
g_{\mu\nu}dx^\mu dx^\nu=(-N^2+\beta_k\beta^k)dt^2+2\beta_kdx^kdt+h_{ij},
\end{align}
where $h_{ij}$ is a Euclidean 3-metric, $N$ is the lapse function, and $\beta$ is the shift vector. Through the Einstein--Hilbert action, one can define a canonical momentum $\pi^{ij}$ associated to the Euclidean metric $h_{ij}$. The Gauss--Codazzi constraint equations on the embedding of a spacelike hypersurface at a constant time can be translated into a Hamiltonian constraint on the canonical variables. When canonically quantized, this constraint gives rise to the Wheeler--deWitt equation.

In the case of an asymptotically AdS spacetime, we have to be a bit more careful, as AdS is not globally hyperbolic, and cannot be foliated in such an easy way. However, by specifying higher derivatives of this asymptotic metric \cite{Kastikainen}, one can get rid of this issue and obtain a well-defined evolution problem. One can then use the Hartle--Hawking state, start with a state in a quantum theory of a time slice of AdS, which could be assimilated to a hyperbolic space described by a Euclidean AdS spacetime compactified in Euclidean time, and use the Wheeler--deWitt equation to choose a Hartle--Hawking state. This gives rise to a static picture \cite{Visser:1989ef}, and the Wheeler--deWitt equation itself does not create any Lorentzian time flow. However, inspired by \cite{Maldacena:2001kr} as well as a modern approach via entanglement entropy \cite{Jafferis:2015del}, we expect the modular flow of the Hartle--Hawking state to be able to describe an extra time dimension.

In such a Hamiltonian context, baby universe formation could arise dynamically, since in the Schrödinger picture, Hamiltonian evolution can allow to go from a state describing a smooth bulk spacetime with a certain topology to a state describing another smooth bulk spacetime with another topology. In particular, it could be possible to evolve between two semiclassical states with different topologies in finite time through some gravitational instantons, as described in \cite{Giddings:1987cg}. In this paper, we set up a precise framework to describe such topology changes directly in terms of the algebra of observables. 

Such a Hamiltonian framework may not be fully replaced by simply considering a Euclidean path integral from Wick-rotated spacetime. There are additional inherently quantum-mechanical effects which are not visible at semiclassical order. Including those will likely enable the path integral to factorize, and these effects will not always be captured in Euclidean gravity. However, for the case of $(1+1)$-dimensions, the Euclidean path integral provides a good approximation, yet in higher dimensions, this does not carry over. In low dimensions, QCD, for example, exhibits normalizable conditions. Even generically, in low dimensions, the Euclidean path integral formalism sometimes provides good enough descriptions for quantum mechanical settings. For higher dimensional operators, the additional quantum features needed are non-normalizable for gravity. The difference between the low and high dimensions is due to the lack of tools that carry over such as the Page curve methods, or more generally speaking, how to get the wormholes. At least for the replica wormholes in three-dimensions, we expect to have a mathematical generalization to occur with a similar mechanism via Kleinian uniformization.

In the specific case of a 4d bulk spacetime, we hope to give a more precise description of such phenomena in future work, by utilizing Kleinian uniformizations to describe the 3d Euclidean geometries. Indeed, the space-like hypersurfaces will then have dimension 3, and their corresponding boundaries will have dimension 2, and hence the boundary CFT's on each time slice will be described by Riemann surfaces. In the case of a spacetime with a single boundary, these Riemann surfaces can be obtained by uniformizations of the Riemann sphere by Schottky groups \cite{Marcolli}, and are topologically dual to handlebodies in AdS$_3$. For each given boundary topology, these handlebodies can then parameterized by the moduli space of Schottky uniformizations \cite{Marcolli}, which corresponds to variations of the bulk geometry. In the case of a bulk with multiple boundaries, one needs a more general class of uniformizations to describe the different bulk handlebodies. We expect that the general theory of Kleinian groups will be the right framework to describe such situations, and will return to this setup in future work.

\section{Diffeomorphism invariance and bulk topology}
\label{sec:difftotopo}

In this section, we give a precise definition of baby universes in terms of wormhole creation and annihilation through connected sums and gluings. We then argue that the requirement of diffeomorphism invariance for a theory of quantum gravity has important consequences in terms of commutativity and associativity of these transformations.

\subsection{Connected sums and gluings}
\label{sec:connetedsums}

The idea of baby universe formation ultimately amounts to having a topology change in the bulk. We now describe the mathematical operations associated to baby universe transformations. The way a baby universe comes into play is through a connected sum with a (maybe empty) component of spacetime. Conversely, the disappearance of a baby universe happens through a gluing of two boundaries of the currently existing spacetime.

\begin{figure}[H]
\vspace{5mm}
\centering
\includegraphics[scale=1.55]{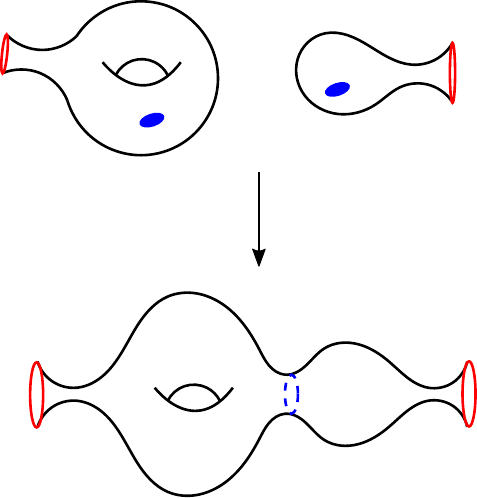}
\vspace{5mm}
\caption{The connected sum operation. The two blue disks are removed from the hyperbolic handlebodies with red conformal boundaries. The handlebodies are then glued together along the holes.}
\label{fig:sum1}
\end{figure}

\begin{figure}[H]
\vspace{8mm}
\centering
\includegraphics[scale=1.55]{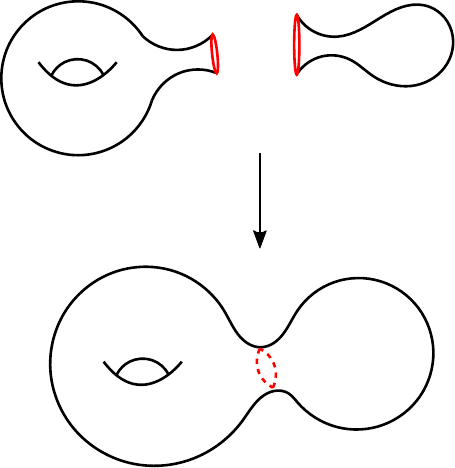}
\vspace{5mm}
\caption{The gluing operation. The two red conformal boundaries are sown together.}
\label{fig:sum2}
\end{figure}

First, the creation of a baby universe can be modelled through a connected sum operation with a manifold with one boundary. A \textit{connected sum} $A\#B$ glues two manifolds $A$ and $B$ together by removing disks from them and gluing them along the disks. An example of a connected sum is depicted in Figure \ref{fig:sum1}. There is essentially two possibilities for our baby universe when it is created: it either gets in a connected sum with a pre-existing piece of the spacetime manifold, or lives on its own. We emphasize that in this latter case, one should not see the independence of the baby universe as nothing happening to the rest of the bulk: physically, a baby universe has still appeared, and it will be the crucial point of our analysis that the underlying bulk theory must know about this. We emphasize this fact by writing that when the baby universe $U$ appears as an isolated island, it is still coupled to the bulk through the connected sum $U\#\emptyset$. 

Similarly, when a baby universe disappears from the bulk theory, it can be seen as happening through a \textit{gluing} of one of the boundaries of the manifold with another manifold with an identical boundary. We will write such an operation $A\sim B$. We demonstrate an example of a gluing operation in Figure \ref{fig:sum2}.

\subsection{Consequences of diffeomorphism invariance}
\label{sec:diffinv}

The operations $\#$ and $\sim$ of connected sums and gluings describe how the creation or annihilation of a baby universe arises. Now, these operations act at the level of the topology of the spacetime manifold, rather than its geometry, and one could imagine a lot of different ways of performing them. In particular, there is a priori no reason to believe that such transformations will be controllable in a nice way in a theory of gravity. In what follows, we will show that in fact, if we believe that our theory of gravity is \textit{diffeomorphism invariant}, the connected sum and gluing operations will inherit a very nice structure. 

Diffeomorphism invariance is a key feature of gravity theories. The easiest way to notice it is to realize that the Einstein--Hilbert action of General Relativity is invariant under diffeomorphism transformations. This diffeomorphism invariance property is believed to hold in the quantum gravity realm: in the semiclassical limit of a large class of quantum gravity theories, including the AdS/CFT correspondence, the Einstein equations (hence diffeomorphism invariance) emerges from the entanglement structure of the theory \cite{Kastikainen,Faulkner:2013ica}. Similarly, in string theory, anomaly cancellation implies that the target space satisfies the Einstein equation, and thus possesses diffeomorphism invariance.

Let us now explore the consequence of diffeomorphism invariance for our connected sum and gluing operations. First, we have the following result from differential topology \cite{Hirsch}.
\medskip
\begin{thm}
The connected sum operation does not depend on where it is performed on a connected manifold, up to diffeomorphism. If two boundaries are identical, the gluing operation does not depend on the boundary it is performed on within the same connected component of a manifold, up to diffeomorphism.
\end{thm}

This result shows that the connected sum of two manifolds can be defined in a unique way in a diffeomorphism invariant theory. In other words, the very large number of possibilities that arise when the connected sum of two manifolds is to be performed is completely gauged out by diffeomorphism invariance. Actually, we have an even stronger result:
\medskip
\begin{thm}
The connected sum and gluing operations are mutually associative and commutative, up to diffeomorphism.
\end{thm}

This result means that the outcome of several connected sum or gluing operations performed in a row will depend neither on the order, nor on the place where the operations are performed. This will have the utmost importance in the rest of our discussion.

\section{The algebra of bulk observables}		\label{sec:bulkalgebra}

So far, our discussion on baby universes has been quite general. In particular, while we specified that creating a baby universe amounts to adding a new boundary to the theory, we have not explicitly implied that our theory of gravity should be described by any specific holographic duality. We now leave the realm of general considerations to focus specifically on what allowing for baby universe creation and annihilation implies in a theory of quantum gravity described by a multi-boundary AdS/CFT correspondence.

In this section, we use the AdS/CFT correspondence to construct the algebra of observables of a bulk theory that allows for the creation and annihilation of baby universes, and therefore a fluctuating number of boundaries. Our starting point will be the Marolf--Wall paradox.\footnote{In order to consider a holographic setup relevant for us, we use the Marolf--Wall perspective for the physically-relevant baby universe algebra, which will be constructed explicitly later. Despite the fact that the commutativity of the baby universe algebra will arise from the properties of topological operations, we still need to justify the existence of the baby universe sector, which will be motivated by the Marolf--Wall model. } 

\subsection{The Marolf--Wall paradox}	\label{sec:MarolfWall}

Here we reformulate the paradox of \cite{Marolf:2012xe} in a different manner, and show that it is relevant to our discussion on baby universes for holographic theories. Let $\mathcal{A}$ be the algebra of observables associated to one boundary CFT. In order for our explanation to be as clear as possible, we will start with the given data of only two identical boundaries, hence the boundary algebra will be of the form $\mathcal{A}\otimes\mathcal{A}$.

The fundamental puzzle Marolf and Wall are pointing out is that only the knowledge of the algebra $\mathcal{A}\otimes\mathcal{A}$ does not make it possible for an observer to differentiate between a situation where this tensor product describes two different theories of quantum bulks with one boundary, or a single theory of a quantum bulk with two boundaries. In the former case, the two bulks will always be independent, whilst in the latter case, wormhole geometries between the two will be allowed in the gravitational path integral. A tempting resolution of the paradox is to say that the answer depends on the state\footnote{The precise definition of a state on an algebra of observables without a reference to a Hilbert space will be given in the next section.} of the system: if it is entangled, then there will be wormhole contributions. However, what Marolf and Wall claim is that this is only an artifact that leaves the real problem unchanged.

More precisely, let us sketch the thought experiment proposed by Marolf and Wall. Let us say that the overall state of the system is known and is a thermofield double state \cite{Israel:1976ur}. Then, the standard argument \cite{VanRaamsdonk:2010pw,Israel:1976ur} shows that the bulk dual when the two boundaries are thought of as being part of the same theory is a wormhole. 

\begin{figure}[H]
\vspace{5mm}
\centering
\includegraphics[scale=1.7]{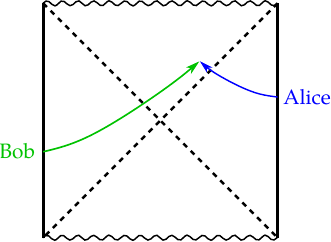}
\vspace{3mm}
\caption{A two-sided semiclassical wormhole corresponding to a thermofield double state. On the right, Alice jumps into the black hole, and one can engineer an operator that acts on the left CFT and creates Bob in such a way that he will meet Alice in the black hole.}
\label{fig:penrose1}
\end{figure}

Let us say Alice jumps into the right side of such a two-sided black hole. An omnipotent agent creates Bob on the left side (by using the left CFT and entanglement wedge reconstruction) and throws him into the black hole, so that Alice and Bob meet inside, and very close to the horizon. Then, there exists an observable $P_1$ in the black hole interior that gives $1$ if Alice and Bob have met and $0$ else, and in this case, $P_1$ will be $1$ with very high probability in the thermofield double state. We demonstrate this scenario in Figure \ref{fig:penrose1}.

\begin{figure}[H]
\vspace{5mm}
\centering
\includegraphics[scale=1.7]{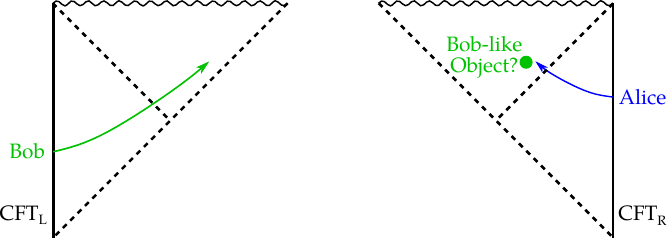}
\vspace{3mm}
\caption{The thermofield double state in the case where the boundaries describe two separate bulks. Assuming approximate entanglement wedge reconstruction close to the horizon, there exists an operator on the right CFT that asks whether Alice meets a ``Bob-like object". This operator commutes with the operator that creates Bob.}
\label{fig:penrose2}
\end{figure}

Now, let us suppose the two CFTs (both left and right CFTs) describe two independent bulks. Then, even in the thermofield double state, if Alice jumps inside a black hole, and we assume that very close to the horizon, the interior can approximately be reconstructed only on the right of the boundary\footnote{Such an assumption requires an approximation theorem for entanglement wedge reconstruction. We come back to this problem in the last section of this paper.}, the encounter with ``a Bob-like object" near the horizon will be described by an operator of the right CFT only, and in particular an observable $P_2$ that measures whether or not Alice finds ``a Bob-like object" in the black hole will be described by an operator of the right CFT. The creation of Bob, on the contrary, will be described by a unitary operator of the left CFT only. 
As such, the conjugate action of this operator on $P_2$ will leave it unchanged, and the probability that Alice sees a Bob inside the black hole will stay close to zero, even in the thermofield double state. This implies that $P_1$ cannot be the same as $P_2$, which would be the case if the two situations were described by the same algebra of observables $\mathcal{A}\otimes\mathcal{A}$! We demonstrate this scenario in Figure \ref{fig:penrose2}.

As a result, according to Marolf--Wall, the two boundary algebras, even in a given state, cannot contain enough information to determine whether or not the theory is seen as a theory of a quantum bulk with two boundaries, or two theories of a quantum bulk with one boundary. In the former case, topologies can interact through entanglement and wormholes can be formed, whereas in the latter case, these interactions are forbidden. The resolution Marolf and Wall propose to the paradox is that the tensor product $\mathcal{A}\otimes\mathcal{A}$ cannot be used to describe the bulk on its own, and that an extra tensor factor $\mathcal{S}$ of ``superselection sectors" must be added. 

The validity of the Marolf--Wall thought experiment is still not firmly established. In particular, as we explain in Section \ref{sec:assumption1}, some arguments in \cite{Marolf:2020xie} tend to show that the reasoning based on thermofield double states in this Lorentzian context is not satisfactory. However, the idea of adding these additional baby universe degrees of freedom, at a state-independent level, is a good starting point for our discussion. Much of this paper will then be devoted to showing that these degrees of freedom may very well disappear once a particular state is chosen.

\subsection{The algebra of baby universes}
\label{sec:AbelianG}

Let us draw the consequences of the Marolf--Wall paradox, and use them to construct an appropriate setup for a bulk theory with baby universes. Consider a reservoir algebra of Euclidean boundary theories 
\begin{align}
\mathcal{A}_{res}:=\bigotimes_b\bigotimes_{n=1}^\infty\mathcal{A}_b^n\otimes\bigotimes_b\bigotimes_{n=1}^\infty\overline{\mathcal{A}_b^n}.
\end{align}
The extra index $b$ accounts for the different boundary structures that one allows in the theory. The algebras $\overline{\mathcal{A}_b^n}$ are seen as conjugates of the corresponding $\mathcal{A}_b^n$. These boundaries are necessary because, to get a $C^\ast$-algebra of observables, we need to know what the conjugate operation is to adding an asymptotic boundary in the dictionary. The interpretation that Marolf--Maxfield give of these conjugate boundaries is that they are added ``in the asymptotic Euclidean past," while the regular boundaries are added ``in the asymptotic Euclidean future."\footnote{We return to the question of the precise definition of these conjugate boundaries in Section \ref{sec:examples}.} 

We want to make a remark that the setup here is a bit different than before, as we are in a Euclidean context. Nevertheless, we will take the viewpoint that the Marolf--Wall thought experiment still teaches us that different AdS/CFT dictionaries are operationally inequivalent, at least at a state-independent	 level.\footnote{We return to this point in Section \ref{sec:assumption1}.} Then, what the Marolf--Wall thought experiment tells us is that the algebra $\mathcal{A}_{res}$ does not contain enough information to determine which AdS/CFT dictionary one should use: which boundaries should be seen as dual to the same bulk theory that is under consideration, and be allowed to form wormholes together in the gravitational path integral, and which should be dual to independent bulk theories. Actually, there is quite a bit of choices for such a dictionary. However, a lot of these choices are diffeomorphism-equivalent. In fact, the properties of connected sums tell us that all configurations where the same number of boundaries and conjugate boundaries of each type (conjugate boundaries counting negatively), is dual to the considered bulk, are diffeomorphism-equivalent.

However, the number of boundaries of each type does make a difference, even after taking into account diffeomorphism (gauge) equivalence. In other words, there is a one-to-one correspondence between the possible numbers of boundaries of the bulk theory and the choices of AdS/CFT dictionary up to diffeomorphism. If one wants to allow for baby universe creation and annihilation, i.e. for changes of AdS/CFT dictionary, there must exist an operator that corresponds to adding a boundary in the bulk gravitational path integral.\footnote{These transformations can be assimilated to the mutually commuting operators that factor out of the Lagrangian in Coleman's initial approach \cite{Coleman:1988tj}, or the operations denoted by $Z[J]$ and $Z^\ast[J]$ in \cite{Marolf:2020xie}.} 

Let us describe the structure of these baby universe operations. The baby universe transformations allow or forbid some boundaries to form wormholes between each other in the gravitational path integral. More precisely, they act in the following way: if the boundaries $\mathcal{A}_b^1$, $\cdots$, $\mathcal{A}_b^n$ and conjugate boundaries $\overline{\mathcal{A}}_b^1$, $\cdots$, $\overline{\mathcal{A}}_b^p$ are interacting through wormholes in the bulk theory, an elementary baby universe transformation of type $b$ adds the boundary $\mathcal{A}_b^{n+1}$, while its conjugate adds $\overline{\mathcal{A}}_b^{p+1}$. Crucially, we saw that the corresponding operations of connected sums can be composed in an associative and commutative way because of diffeomorphism invariance. 

This means that diffeomorphism invariance tells us something extremely strong: due to the properties of the connected sum, the baby universe operations form an Abelian algebra! As we shall see, this commutativity will have a crucial importance in the construction of the baby universe Hilbert space and the formulation of the baby universe hypothesis. Analogous to the framework of algebraic quantum field theory (AQFT) \cite{Haag}, we require the baby universe algebra $\mathcal{A}_{baby}$ to be a $C^\ast$-algebra, i.e. it possesses an involution operation and is closed for a given norm. Note that in this algebra, because adding a conjugate boundary can be seen as a topological operation that commutes with the connected sum, the operator associated with the addition of an asymptotic boundary will commute with its adjoint, and hence be a \textit{normal} operator.

The Marolf--Wall paradox shows that there is no way to determine whether or not a boundary has been allowed to form bulk wormholes and play the role of a baby universe just by looking at the boundary reservoir algebra $\mathcal{A}_{res}$. In particular, it follows that the baby universe transformations will act on it trivially. Since the baby universe operations have a nontrivial action on the bulk observables, we are bound to conclude that the algebra of observables of bulk quantum gravity $\mathcal{A}_{QG}$ satisfies
\begin{align}
\mathcal{A}_{QG}\neq\mathcal{A}_{res}.
\end{align}

This means that $\mathcal{A}_{baby}$ has to be added to the algebra of observables of bulk quantum gravity in order to fully describe the bulk in a holographic setup with a fluctuating number of asymptotic boundaries. In the language of quantum mechanics, the description of the addition of the associated ``extra" degrees of freedom is a tensor product operation. Hence we conclude 
\begin{align}
\mathcal{A}_{QG}=\mathcal{A}_{res}\otimes\mathcal{A}_{baby},
\end{align}
where $\mathcal{A}_{baby}$ is an Abelian $C^\ast$-algebra.

\section{The Hilbert space of baby universes}
\label{sec:hilbertBU}

The analysis of the last section culminated in the construction of the algebra of observables of bulk quantum gravity theory in a holographic context
\begin{align}
\mathcal{A}_{QG}:=\mathcal{A}_{res}\otimes \mathcal{A}_{baby},
\end{align}
where $\mathcal{A}_{res}$ is the reservoir algebra and $\mathcal{A}_{baby}$ is the abelian $C^\ast$-algebra of baby universes. This baby universe algebra corresponds to the baby universe operators, which, as the Marolf--Wall thought experiment teaches us, cannot be contained in the boundary reservoir.

We now want to construct the Hilbert space of baby universes out of this new algebra. Such a construction requires an extra input: a state on the algebra $\mathcal{A}_{QG}$, which is the equivalent of the gravitational path integral in our setup. In this section, we define a gravitational path integral as a state on $\mathcal{A}_{QG}$ and mimic Marolf--Maxfield's construction of the baby universe Hilbert space through a standard technique in $C^\ast$-algebra theory and the GNS construction. It is important to note that, in what follows, the construction of the baby universe Hilbert space will only depend on the commutative nature of the algebra of baby universes; hence, it will be valid in a more general context than holography.

\subsection{The gravitational path integral and the GNS construction}
\label{sec:pathintegral}

In a physical Hilbert space, there is always a reference state, which corresponds to the vacuum of the theory: typical examples include the QFT vacuum or a thermal state. This reference state actually comes about naturally when one thinks of the Hilbert space as not being the fundamental object of the quantum theory, but as being derived from its algebra of observables. Indeed, in order to derive a Hilbert space from a $C^\ast$-algebra, one has to perform the so-called GNS construction with respect to an abstract state on the algebra of observables, which is defined as an expectation value functional. In a theory of quantum gravity in the bulk, this role is played by the gravitational path integral, which associates to each bulk operator its expectation value by ``summing over geometries."

Formally, the gravitational path integral $\omega$ is a state on the algebra of observables $\mathcal{A}_{QG}$, i.e. it is a continuous linear functional that satisfies 
\begin{align}
\omega(Id)=Id,
\end{align}
and for $A\in\mathcal{A}_{QG}$, 
\begin{align}
\omega(A^\ast A)\geq 0.
\end{align}
There are many possible choices of such a state, and the space of states can be very big in general. That is the reason why picking a state (in our case, a gravitational path integral) is an important piece of information that will change a lot of the physics. Now, say that that we are given the algebra $\mathcal{A}_{QG}$ and a state (i.e. gravitational path integral) $\omega$ on it. This is enough data to construct a Hilbert space for the theory, through the GNS construction, which we proceed to explain. 

The GNS representation constructs the Hilbert space directly out of the algebra: to each element $A\in\mathcal{A}_{QG}$, we formally associate a vector $\ket{A}$. Now, we consider the linear span of such vectors, and we define an inner product through the gravitational path integral $\omega$ (in an intuitive manner) as
\begin{align}
\braket{A,B}:=\omega(A^\ast B).
\end{align}
The property 
\begin{align}
\omega(A^\ast A)\geq 0
\end{align}
ensures that this inner product is positive, however it need not be positive definite. We then need to quotient out the null states, as similarly performed in \cite{Marolf:2020xie}. Define 
\begin{align}
\mathcal{I}:=\left\{A\in\mathcal{A}_{QG}, \omega(A^\ast A)=0 \right\} .
\end{align}
The GNS Hilbert space is then given by the quotient 
\begin{align}
\mathcal{H}:=\mathcal{A}_{QG}/\mathcal{I} \, ,
\end{align}
equipped with the induced inner product, which is now positive definite. The requirement that $\mathcal{A}_{QG}$ is a $C^\ast$-algebra is enough to prove that $\mathcal{H}$ is indeed a Hilbert space \cite{Takesaki}. In particular, the vector $\ket{\Omega}$ that arises from the equivalence class of the identity in this quotient satisfies
\begin{align}
\bra{\Omega}A\ket{\Omega}=\omega(A),
\end{align}
for any $A\in\mathcal{A_{QG}}$, and therefore should be seen as the vacuum state of the Hilbert space.  

The claim of Marolf--Maxfield is that in the case of baby universes, the quotient operation will actually dramatically collapse the Hilbert space. The Baby universe hypothesis of McNamara--Vafa is even stronger, as it states that the baby universe part of the Hilbert space is of dimension one. In what follows, we show that these claims actually follow very naturally from the fact that the baby universe $C^\ast$-algebra is commutative: its GNS theory will then show similar collapse features.

\subsection{Reduction to an Abelian GNS construction}

Let us now try to apply the GNS construction to our case. Recall that the algebra of quantum gravity observables $\mathcal{A}_{QG}$ takes the form:
\begin{align}
\mathcal{A}_{QG}:=\mathcal{A}_{res}\otimes \mathcal{A}_{baby},
\end{align}
where $\mathcal{A}_{baby}$ is the \textit{commutative} $C^\ast$-algebra of baby universes. In order to construct the Hilbert space, we need the extra input of the abstract state $\ket{\omega}$ (expectation value functional) from which the gravitational path integral arises.

In particular, $\omega$ restricts to $\mathcal{A}_{baby}$, and one can take the GNS representation of $\mathcal{A}_{baby}$ with respect to $\omega$ to construct the baby universe Hilbert space. Here comes the crucial point: recall that the algebra of baby universe observables $\mathcal{A}_{baby}$ is \textit{commutative}. The problem of understanding the structure of the baby universe Hilbert space boils down to understanding the GNS theory in the Abelian case. Fortunately enough, in this case, a lot can be said, especially about the baby universe hypothesis.

\section{Refined baby universe hypothesis}
\label{sec:BUH}
We now consider the baby universe hypothesis formulated  by McNamara--Vafa in the context of the GNS representation and our formalism. We refine the baby universe hypothesis utilizing the GNS representation and the formalism we have built in Section \ref{sec:purity}, and find the necessary and sufficient mathematical condition for the baby universe hypothesis to be valid in Section \ref{sec:newBUH}. We further study its physical validity and derive a theorem with three physical conditions for the baby universe hypothesis to hold.

\subsection{The baby universe hypothesis as a purity requirement}
\label{sec:purity}

We first recall the baby universe hypothesis by McNamara--Vafa.
\medskip
\begin{hypo}[Baby universe hypothesis \cite{McNamara:2020uza}]
Let $\calh_{BU}$ be the Hilbert space of baby universes in a unitary theory of quantum gravity in $d>3$ spacetime dimensions. Then we have $\mathrm{dim}\calh_{BU}=1$.
\end{hypo} 

It follows from the previous discussion that in our framework, the baby universe hypothesis transforms into the following statement:
\medskip
\begin{hypo}[Baby universe hypothesis, reformulation 1]
Let $\omega$ be the gravitational path integral state on the algebra of observables of quantum gravity $\mathcal{A}_{QG}$. A GNS representation of the baby universe algebra $\mathcal{A}_{baby}$ arising from $\omega$ has dimension one.
\end{hypo}

In other words, in order to determine when the baby universe hypothesis is true, one needs to know when a state on a commutative $C^\ast$-algebra gives rise to a the GNS representation of dimension one. The answer is beautiful and simple utilizing the following theorem, which is reformulated from \cite[Corollary 3.6]{Stormer}.\footnote{Stormer formulates this result in terms of pure states. The rest of our subsection clarifies the equivalence between the two formulations.}
\medskip
\begin{thm}
Let $\mathcal{A}$ be a commutative $C^\ast$-algebra and $\omega$ be a state on $\mathcal{A}$. The GNS representation of $\mathcal{A}$ has dimension one if and only if it is irreducible.
\end{thm}

We note that this theorem is completely false if $\mathcal{A}$ is not required to be commutative: there could very well exist representations of $\mathcal{A}$ that are irreducible but have a dimension much higher than one! As a result, the fact that the GNS construction collapses so much of the Hilbert space in the baby universe case can be traced back to the fact that the baby universe algebra $\mathcal{A}_{baby}$ is commutative, which itself, as we saw, comes from diffeomorphism invariance. In other words, diffeomorphism invariance forces the baby universe hypothesis to be true in any irreducible GNS representation of the algebra of baby universes! 

We have shown the following reformulation of the baby universe hypothesis:
\medskip
\begin{hypo}[Baby universe hypothesis, reformulation 2]
Let $\omega$ be the gravitational path integral state on the algebra of observables of quantum gravity $\mathcal{A}_{QG}$. The GNS representation of the baby universe algebra $\mathcal{A}_{baby}$ arising from the restriction of $\omega$ to $\mathcal{A}_{baby}$ is irreducible.
\end{hypo}

Now, what does it mean for a state on a $C^\ast$-algebra to have an irreducible GNS representation? It turns out that the answer to this question is once again very simple. Recall that the space of states on a $C^\ast$-algebra is the space of expectation value functionals on this algebra. The space of states is a convex space, as a convex combination of expectation value functionals is an expectation value functional. As a result, this space has extremal points, i.e. points which can not be expressed as a convex combination of two other ones. This is the abstract definition of a \textit{pure state}, as a state that cannot be expressed as a statistical mixture of other states. The following theorem is a well-known fact in $C^\ast$-algebra theory:
\medskip
\begin{thm}
Let $\mathcal{A}$ be a $C^\ast$-algebra, and $\omega$ be a state on $\mathcal{A}$. The GNS representation of $\omega$ is irreducible if and only if $\omega$ is pure.
\end{thm}

This theorem, put together with our second reformulation of the baby universe hypothesis, justifies our final reformulation:
\medskip
\begin{hypo}[Baby universe hypothesis, reformulation 3]
The gravitational path integral arises from a state $\omega$ whose restriction to the baby universe algebra is pure.
\label{hypo:BUTiff}
\end{hypo}

Note that here, a clear bridge can be drawn between the baby universe hypothesis and the question whether the baby universe theory should be described as an ensemble of theories. Indeed, the space of states on a $C^\ast$-algebra is a bit more than a convex space: it is a Choquet simplex. This means that every state $\omega$ on the baby universe algebra $\mathcal{A}_{baby}$ can be written as a formal integral over the space of pure states:
\begin{align}
\omega=\int_{\varphi\;\mathrm{pure}}\varphi d\mu(\varphi),
\end{align}
for a certain measure $\mu$ on the space of pure states. In terms of the GNS representations, this will be translated into a direct integral decomposition of the GNS Hilbert space:
\begin{align}
\mathcal{H}_{\omega}=\int^{\bigoplus}\mathcal{H}_{\varphi}d\mu(\varphi),
\end{align}
where the $\mathcal{H}_{\varphi}$ are irreducible GNS representations (in our case, dimension one Hilbert spaces). Each member of this direct integral decomposition, or equivalently of the decomposition of $\omega$, determines one member of the ensemble of theories that is averaged over. This ensemble contains one element if and only if $\omega$ is pure, which is equivalent to the baby universe hypothesis.

Therefore, the baby universe hypothesis is valid if and only if the theory of baby universes is not described by an ensemble. However, we emphasize that the gravitational path integral on the whole algebra $\mathcal{A}_{QG}$ corresponding to a pure state does not exactly correspond to the baby universe hypothesis. Indeed, the baby universe hypothesis assumes that the \textit{restriction} of this state to the algebra of baby universes $\mathcal{A}_{baby}$ is pure. In particular, this would imply that $\omega$ can be written as\footnote{For a proof, see \cite[Lemma 4.11]{Takesaki}.} 
\begin{align}
\omega=\omega_{res}\otimes\omega_{baby}.
\end{align}
In other words, a necessary extra assumption in order for the baby universe hypothesis to be satisfied is that the gravitational path integral factorizes, in the form:
\begin{align}
Z_{QG}=\int_{CFT}D[\phi]e^{-S[\phi]}\int_{baby}D[b]e^{-S[b]}.
\end{align}
This implies that the formation of the baby universes cannot be influenced by the boundary observables, which we will explain in detail in Section \ref{sec:assumption3}.

We note that throughout this discussion, no assumption specific to holography was required. The dimension of any Hilbert space constructed from a GNS representation of any commutative algebra of observables will follow the same rules.

\subsection{Baby universe theorems}
\label{sec:newBUH}

Based on our analysis in Section \ref{sec:purity}, we first state Theorem \ref{thm:BUpure} that formulates a condition of purity that is the only required criterion for the baby universe hypothesis to be valid. 
Investigating this theorem in the context of physical setups, we interpret this purity requirement into sufficient conditions on a gravitational path integral state. This involves satisfying the assumptions as listed in Theorem \ref{thm:BU}. We will discuss the physical validity of these assumptions in Section \ref{sec:assumptions}.
\medskip
\begin{thm:BUpure}
Let $\mathcal{A}_{QG}$ be the algebra of observables of quantum gravity, and suppose that $\mathcal{A}_{QG}=\mathcal{A}_{res}\otimes \mathcal{A}_{baby}$, where $\mathcal{A}_{baby}$ is Abelian. Let $\omega$ be a state on $\mathcal{A}_{QG}$. The GNS representation of $\mathcal{A}_{baby}$ induced by $\omega$ has dimension 1 if and only if the restriction of $\omega$ to $\mathcal{A}_{baby}$ is pure.
\end{thm:BUpure}
\medskip 
\begin{thm:BU}
Let $\mathcal{A}_{QG}$ be the algebra of observables of bulk quantum gravity, and $\omega$ be a gravitational path integral state on $\mathcal{A}_{QG}$. Assume that

\begin{enumerate}
\item $\mathcal{A}_{QG}=\mathcal{A}_{res}\otimes \mathcal{A}_{baby}$ for $\mathcal{A}_{baby}$ an Abelian $C^\ast$-algebra, 
\item $\omega$ is a pure state on $\mathcal{A}_{QG}$,
\item $\omega$ can be factorized as $\omega=\omega_{res}\otimes\omega_{baby}$.
\end{enumerate}
Then, the GNS representation of $\mathcal{A}_{baby}$ is one-dimensional.
\end{thm:BU}

Recall that the first assumption amounts to the Marolf--Wall proposal adapted to our setup, which will be discussed in Section \ref{sec:assumption1}. The second assumption states that the theory is not described by an ensemble, but by a single Hamiltonian. In particular, we showed earlier that this amounted to the GNS representation of $\omega$ being irreducible. We discuss the validity of the second assumption in Section \ref{sec:assumption2}. Finally, the third assumption requires the factorization of the gravitational path integral between the baby universe observables and the other ones. The third assumption is discussed in Section \ref{sec:assumption3}.

\section{``Miraculous cancellations" and $\alpha$-parameters}
\label{sec:alpha}
 We now turn to investigating the implications of our formalism. In this section, we utilize the Gelfand isomorphism for Abelian $C^\ast$-algebras to explain the ``miraculous cancellations" of the bulk topologies as a natural consequence of character theory. We then link our construction to the $\alpha$-parameters of baby universes.

One question remains unsolved: if the baby universe hypothesis is true, and the gravitational path integral arises from a state whose restriction to the algebra of baby universe observables $\mathcal{A}_{baby}$ is pure, then, what should such a state look like? Once again, the theory of Abelian $C^\ast$-algebras comes to the rescue:
\medskip
\begin{thm}[Proposition 4.4.1, \cite{Kadison}]
A state $\omega$ on an Abelian $C^\ast$-algebra $\mathcal{A}$ is pure if and only if it is multiplicative, i.e., for $A$ and $B$ in $\mathcal{A}$: 
\begin{align*}
\omega(AB)=\omega(A)\omega(B).
\end{align*}
Such a state is called a \textit{character} of $\mathcal{A}$.
\end{thm}

This is exactly equivalent to the claim of McNamara--Vafa that when the baby universe Hilbert space has dimension one, the correlation functions of the effects of adding two boundaries factorize. They argue that this requires ``miraculous cancellations" of contributions of different topologies in the bulk. Here, we attempt to show a little bit more explicitly how this ``miracle" can happen.

The crucial ingredient for the following discussion is the following theorem: 

\begin{thm}[\cite{Takesaki}]
Let $\mathcal{A}$ be an Abelian $C^\ast$-algebra. $\mathcal{A}$ is isomorphic to the algebra $C_0(X)$ of complex-valued continuous functions on $X$ vanishing at infinity, where $X$ is a locally compact Hausdorff space. $X$ is compact if and only if $\mathcal{A}$ contains a unit element, and is called the Gelfand space of $\mathcal{A}$.
\end{thm}

In other words, any Abelian $C^\ast$-algebra can be assimilated to the algebra of complex valued continuous functions on a locally compact space (the involution corresponding to complex conjugation). The interest of the Gelfand space is that it provides a very simple description of the states and characters of $\mathcal{A}$:

\begin{thm}[\cite{Takesaki}]
Let $\mathcal{A}=C_0(X)$ be an Abelian $C^\ast$-algebra. States on $\mathcal{A}$ are Radon probability measures on $X$, and characters on $\mathcal{A}$ correspond to Dirac measures on $X$.
\end{thm}

This means that the space of pure states, also called $\alpha$-parameters in \cite{Marolf:2020xie} on $\mathcal{A}$ corresponds to evaluation functionals on $X$, which are directly isomorphic to $X$! As a result, the space of characters on $\mathcal{A}$ which realize the ``miraculous cancellations" in the gravitational path integral has a very simple description in terms of the structure of $\mathcal{A}$: it corresponds to its Gelfand space.

It is then fascinating to us to see that whilst the space of states on an Abelian $C^\ast$-algebra is extremely large, the moduli space of states which satisfy the baby universe hypothesis (or $\alpha$-states) is reduced to a locally compact space, which is even compact in the unital case. Out of the many possible path integrals, the baby universe hypothesis singles out a very small amount of chosen ones, which will cancel the observables in the algebra of baby universes enough that they all collapse into a trivial tensor factor at the level of the Hilbert space. Just like a swampland condition, the baby universe hypothesis seems to invalidate almost every possible choice of gravitational path integral.

As demonstrated, the concept of ``miraculous cancellations" is deeply rooted in utilizing the Gelfand isomorphism. We discuss the physical significance of the Gelfand space in more detail in Section \ref{sec:summary}.

\section{Two examples}
\label{sec:examples}

In this section, we illustrate the previous results by applying them to two different settings. In \ref{sec:topo}, we first consider in the context of a topological theory considered in \cite{Marolf:2020xie} in interpret in the context of our formalism. In \ref{sec:group}, we apply our construction when the action of adding a boundary is invertible, which allows us to derive a precise analogy between our framework and the vacuum structure of QCD.

\subsection{A topological theory} \label{sec:topo}

In \cite{Marolf:2020xie}, Marolf--Maxfield proposed a topological theory of surfaces in order to probe the baby universe construction. In this very simple toy model, only one boundary structure is allowed and the operator that inserts a boundary in the gravitational path integral is called $Z$. This $Z$ is defined to be Hermitian, so that 
\begin{align}
Z^\dagger=Z,
\end{align}
and adding a conjugate boundary is indistinguishable from adding a boundary. The idea is then to identify $Z$ with the real variable function 
\begin{align}
f(x)=x,
\end{align}
which is invariant under complex conjugation, and to look at the algebra it generates, together with the constant function equal to 1. Then, a state $\omega$ that represents the gravitational path integral is defined on this algebra by the topological theory, and it is found that the moments $\omega(Z^n)$ are the ones of a Poisson law. This uniquely determines the measure $\omega$ as a Poisson measure, as the Poisson law's generating function has a positive convergence radius. As the Poisson law has a density that vanishes everywhere except on the positive integers, it is then argued that the GNS representation will have an extremely collapsing effect, as all functions that vanish on the positive integers will be cancelled by the state $\omega$. This can be seen a bit more formally in that the naive direct integral
\begin{align}
\omega=\int_{\mathbb{R}}\varphi d\mu(\varphi),
\end{align}
is only a discrete sum
\begin{align}
\omega=\sum_{n=0}^\infty\delta_n P(Z=n).
\end{align}
This will result in a discrete splitting of the corresponding direct integral of baby universe Hilbert spaces. In a way, Marolf--Maxfield show that their topological theory was ``almost" a pure state on the real line: before the GNS representation, there was a continuum of $\alpha$-states, but only a countable number of them survive in the baby universe Hilbert space. 

Note that there is a technical caveat in this construction: the function $f(x)=x$ and the constant function 1 are not vanishing at infinity and hence do not satisfy the hypotheses of Gelfand duality. However, an easy fix is to perform the same GNS construction on the polynomial algebra generated by $f$, without requiring a Hilbert space structure, but just a pre-Hilbert space structure. This pre-Hilbert space structure is well-defined since fortunately enough, all squares of polynomials have a finite expectation value when integrated against the Poisson density. Then the completion can be performed directly at the level of the Hilbert space.

In \cite{Marolf:2020xie}, the baby universe algebra is enriched by, for example, adding pairs of end of the world branes that represent CFT degrees of freedom, linked by boundary segments. As the resulting observables also commute with the other baby universe observables, the resulting algebra is still commutative. It follows that the same reasoning can carry over. In that case, the new (generalized) Gelfand space is the product of $\mathbb{R}$ and of a space of matrices, and the gravitational path integral state with respect to which the GNS construction is performed is now a joint probability density between different random variables with support on some directions of the Gelfand space, for example, a complex Wishart distribution.

One can further enhance the model by allowing one to slice some asymptotic boundaries and add single end of the world branes or boundary halves. However, the presence of different states on the end of the world branes requires these operations to no longer commute, hence the considered algebra is no longer Abelian. Regardless, \cite{Marolf:2020xie} argue that in this extended algebra, the GNS representation is once again collapsing the Hilbert space significantly. It identifies an entangling operation associated to end of the world branes with the Hartle--Hawking preparation of an Einstein--Rosen bridge, hence casting doubt on the original Marolf--Wall thought experiment. We return to this point in Section \ref{sec:assumption1}.

\subsection{The group case: analogy with QCD and $\theta$-vacua} \label{sec:group}

Another application of our framework is to consider the case where the conjugate boundary operator is the inverse of the boundary operator, so that 
\begin{align}
Z^\dagger=Z^{-1}.
\end{align}
This can be achieved by seeing the conjugate boundary operation as a gluing against its corresponding boundary. The gravitational path integral is then performed over topologies with one less boundary than before. As the operation of adding a boundary therefore has an inverse, the algebra generated by $Z$ will inherit a group structure, and will be a group $C^\ast$-algebra. In this case, a strong analogy with the topological structure of QCD can be made.

A group $C^\ast$-algebra is defined in the following way. Consider a group $G$. One can formally consider the algebra of formal linear combinations of elements on the group, 
\begin{align}
\mathbb{C}[G]:=\left\{\sum_{i=1}^n\lambda_i g_i, \lambda_i\in\mathbb{C}, g_i\in G\right\},
\end{align}
with the multiplication law given by the group multiplication. This set is called the group algebra of $G$. In particular, for the commutative free group with $k$ generators $\mathbb{Z}^k$, which corresponds to the baby universe group in the case where $k$ boundary structures are allowed, $\mathbb{C}[\mathbb{Z}^k]$ will be isomorphic to the algebra of polynomials with $k$ variables and positive and negative powers. In the simplest case where only one boundary structure is allowed in the reservoir, the group algebra $\mathbb{C}[\mathbb{Z}]$ will be isomorphic to the space of Laurent series in one variable.

We have almost constructed the group $C^\ast$-algebra. However, we must bear in mind that in the end, we want to construct the Hilbert space of the theory, on which we will need to put a norm. In order for the additional topological structure of the Hilbert space to be well-defined, we need to do a little more with the group algebra, and endow it with an involution $^*$ (corresponding to taking the group inverse of each group element and the complex conjugate of each scalar) and a norm structure that satisfies the identity 
\begin{align}
\|x^*x\|=\|x\|^2,
\end{align}
so that it becomes a $C^\ast$-algebra.

It turns out that representation theory arguments \cite{Brown,Putnam} show that there exists a unique way to construct such a norm on a group algebra. Then, the completion of the group algebra with respect to this canonical norm gives a group $C^\ast$-algebra, which is denoted as $C^\ast(G)$ and contains enough structure to construct Hilbert spaces. If $G$ is Abelian, it is straightforward to see that the corresponding group $C^\ast$-algebra will also be Abelian, and this fact will be crucial for the last part of our analysis.

Then, the Gelfand dual of the group $C^\ast$-algebra $C^\ast(G)$ is the space of characters of $G$, which is in that case called the Pontryagin dual $\hat{G}$. In the case where $k$ boundary structures are allowed, the baby universe group $G$ is the free Abelian group with $k$ generators $\mathbb{Z}^k$. As characters are multiplicative, their values are entirely determined by the ones they take on the generators of $G$. As a result, the space of pure states on the baby universe algebra when $k$ boundaries are allowed is isomorphic to $k$ copies of a circle, i.e. a $k$-torus. In particular, if only one boundary is allowed, then the space of pure states is isomorphic to the unit circle.

We can now reinterpret the group $G$ of baby universe transformations as a global symmetry group for our theory, and draw an analogy with the $\theta$ angle in QCD.

Suppose that the restriction of the gravitational path integral to the baby universe algebra is a pure state $\omega$. Then, we observed that $\omega$ is a character of the Abelian group of baby universes 
\begin{align}
G=\mathbb{Z}^k ,
\end{align}
where $k$ is the number of allowed boundary structures. Take an observable of quantum gravity $O\in\mathcal{A}_{QG}$. This observable $O$ has the form 
\begin{align}
O=\sum_p a_p A_p\otimes B_p,
\end{align}
where $A_p\in\mathcal{A}_{res}$ and $B_p\in\mathcal{A}_{baby}$. After the action of the baby universe element 
\begin{align}
g=\sum_{i=1}^kn_i g_i,
\end{align}
with $n_i\in\mathbb{Z}$ and $g_i$ being the $k$ independent generators of $G$, $O$ becomes 
\begin{align}
O_g=\sum_p a_p A_p\otimes gB_p.
\end{align}
If $\omega$ is a state that satisfies the baby universe hypothesis, we then get
\begin{align}
\omega(O_g)=\sum_p\omega(A_p)\omega(g)\omega(B_p)=e^{i\sum_{i=1}^kn_i\alpha_i}\sum_p\omega(A_p)\omega(B_p),
\end{align}
where $\omega(g_i)=e^{i\alpha_i}$.
The action of a baby universe transformation therefore adds a phase to the action for the gravitational path integral on an observable. In a way, it preserves the gravitational up to a phase. This phase is parameterized by the real numbers $\alpha_i$ which span the Pontryagin dual 
\begin{align}
\hat{G}=U(1)^k, 
\end{align}
and these real numbers are the analogues of the $\alpha$-parameters of \cite{Preskill:1988na}.

One can then interpret the group of baby universe transformations $G$ as a global symmetry group, and the associated $\alpha$-parameters as topological charges. This is very close to the context of non-Abelian gauge theories, at least in Euclidean signature \cite{Forkel:2000sq}, where similar phase phenomena appear due to topological obstructions. Indeed, the nontriviality of the homotopy group $\pi_3(G)=\mathbb{Z}$, where $G$ is any compact connected simple Lie gauge group (for example the one of QCD), leads to similar phases in a class of path integrals. These phase terms are of the form $e^{in\theta}$, where $n$ is the winding number of the gauge transformation and $\theta$ is the vacuum angle, which plays the exact same role as the $\alpha$ parameters, and labels the superselection sectors of the theory, in the same way as $\alpha$ states give rise to irreducible GNS representations in our case. Here, the analog of the winding number is the number of boundaries (conjugate boundaries counting negatively) added by the gauge transformation.

Note that this framework is reminiscent of the global gauge group construction in algebraic quantum field theory \cite{Haag}. If a quantum field theory is described in terms of its net of algebras of observables, its irreducible representations are labelled by the Pontryagin dual of a group, called the \textit{global gauge group} of the theory. Here, the analogy would be that the global symmetry group is the group $G$ of baby universe transformations, and the irreducible representations are labeled by its Pontryagin dual $\hat{G}$ of $\alpha$-parameters. This group of $\alpha$-parameters can be interpreted as a group of topological charges, or equivalently, of charge-carrying transformations. 

However, in our context, one has to be careful with such a gauge group denomination. Indeed, in AQFT, the global gauge group acts trivially on the (local net of) observables. This encodes a global version of the general notion of a local gauge group $G$, which acts by gauge transformations on the space of field configurations and can be formulated as a structure group of $G$-principal bundles in the configuration space of a (classical) gauge theory. Rather than encoding an actual symmetry, this global gauge group should then rather be thought of as encoding a redundancy.\footnote{Local gauge redundancy is much harder to encode in the framework of AQFT, although some recent progress has been made in \cite{Rejzner:2020bsc} through the BV formalism.} Here, we have more than a gauge redundancy, as we saw that the gravitational path integral picks up a phase when a baby universe transformation is applied, due to the topological charge that is carried. For this reason, we emphasize that our global symmetry group cannot really be interpreted as a gauge group, but instead encodes a global, topological symmetry of our theory.

\section{Probing the assumptions of the baby universe theorem}
\label{sec:assumptions}

In this section, we probe the three assumptions of Theorem \ref{thm:BU}, which states that when they are put together, they lead to the validity of the baby universe hypothesis. We note that the first assumption is, in fact, a more general condition (in the context of the setup of holographic theories), underpinning the baby universe construction of Marolf--Maxfield. 

\subsection{Assumption 1: the Marolf--Wall paradox and approximate entanglement wedge reconstruction}
\label{sec:assumption1}

In the context of AdS/CFT, the first assumption amounts to accepting the conclusions of the Marolf--Wall thought experiment, which we rehearsed in \ref{sec:MarolfWall}. Indeed, we showed that if the different choices of AdS/CFT dictionaries were operationally distinct, we could construct the algebra of observables of the bulk by taking the tensor product of the boundary reservoir algebra with $\mathcal{A}_{baby}$, the $C^\ast$-algebra of baby universes. However, we note that the commutativity of the obtained algebra of baby universes then only relies on topological properties of connected sums, which we expect to still hold in more general contexts.

In order to show that dictionaries with a different number of bulk theories interacting through wormholes were operationally distinct, Marolf and Wall had to make an essential assumption in their thought experiment: the near-horizon physics in the black hole interior were supposed to be approximately reconstructed from Alice's side of the boundary. If this assumption is not here, the operator asking whether Alice finds a Bob-like object near the horizon of the black hole will not be defined on Alice's boundary only, which will invalidate the rest of the thought experiment.

As a result, in order to fully trust the conclusions of the Marolf--Wall thought experiment, we need to know if such an approximate entanglement wedge reconstruction result could hold near the horizon of a single boundary black hole in a given state. For a single boundary in a thermal state, the exact entanglement wedge stops at the black hole horizon, and nothing a priori tells us that such a reconstruction will be possible, even very close to the horizon. In fact, \cite{Junge:2015lmb} show that such results could exist, but are often very nontrivial and state dependent; a few years later, a similar result is shown in \cite{Hayden:2018khn}. If entanglement wedge reconstruction were to fail drastically even very close to the black hole horizon, the Marolf--Wall thought experiment would become invalid, and the physics of the Marolf--Maxfield construction would become unclear. 

In particular, quoting the result of \cite{Marolf:2020xie} in the language of GNS representations, it suggests that in the Lorentzian case, the GNS construction of an algebra of baby universes extended with end of the world branes and half boundaries collapses the two-sided Einstein--Rosen bridge, prepared by a Hartle--Hawking protocol over a half-space, and a state prepared by entangling two boundary cuts onto the same state. This would imply that, in the particular setup of the original Marolf--Wall proposal, the two thermofield double states of the thought experiment are in the same GNS equivalence class, therefore casting doubt on the initial reasoning. In our Euclidean case, it is less clear whether a Marolf--Wall-like thought experiment could rigorously imply the necessity for superselection sectors. We leave such investigations on approximate entanglement wedge reconstruction to the future, and remain agnostic about the range of validity of the Marolf--Wall thought experiment for the time being. If some of these sectors can actually be proven to be indistinguishable in a realistic theory (which amounts to the baby universe hypothesis), it does not hurt to postulate their existence prior to the GNS representation, as their contribution will then vanish at the level of Hilbert spaces. In fact, the collapse of the GNS representation constitutes a proof of this equivalence between sectors.

\subsection{Assumption 2: ensembles in quantum gravity}
\label{sec:assumption2}

As we discussed in Section \ref{sec:purity}, assuming that the gravitational path integral $\omega$ is pure on the algebra of observables of quantum gravity amounts to imposing that it cannot be decomposed into a convex combination of other states: equivalently, it describes a single theory and not an ensemble, i.e. if the integral 
\begin{align}
\omega=\int_{\varphi\;\mathrm{pure}}\varphi d\mu(\varphi)
\end{align}
is reduced to a single summand, then the GNS representation is irreducible. Recent works show that there actually exist some theories of quantum gravity in two dimensions which are dual to ensembles \cite{KitaevTalk15}. In particular, Jackiw--Teitelboim gravity has recently been reinterpreted in terms of the Sachdev--Ye--Kitaev model \cite{Maldacena:2016hyu,Kitaev:2017awl,KitaevTalk14} and matrix integrals \cite{Saad:2019lba,Stanford:2019vob}. We also constructed a general setup for the semiclassical limit of holographic models out of the GNS representations of thermal ensembles in \cite{MonicaElliott}.

In \cite{McNamara:2020uza}, McNamara--Vafa use swampland arguments to argue that such examples are specific to low dimensions. Such swampland considerations might lead to a proof that a rigorous theory of quantum gravity in high enough dimensions cannot be described by an ensemble. For example, the original AdS/CFT correspondence \cite{Maldacena:1997re} does not not use ensembles. However, even if it turned out to be correct, we emphasize that the second assumption on its own does not ensure that the baby universe hypothesis is correct. It is important to note that the third assumption is the crucial one (though not on its own) that ensures that the restriction of the gravitational path integral to the baby universe algebra is pure.

\subsection{Assumption 3: topology changes from matter interactions}
\label{sec:assumption3}

The third assumption required for the baby universe hypothesis to be correct is, in our opinion, the most problematic one. Indeed, assuming that the gravitational path integral factorizes in the form
\begin{align}
\omega=\omega_{res}\otimes\omega_{baby},
\end{align}
or equivalently, 
\begin{align}
Z_{QG}=\int_{CFT}D[\phi]e^{-S[\phi]}\int_{baby}D[b]e^{-S[b]},
\end{align}
implies that no interaction is allowed between the regular bulk observables that can be described from the boundary once the AdS/CFT dictionary is chosen, and the baby universe observables. 

This would mean that baby universe formation cannot be influenced in any way by any traditional bulk process coming from the traditional AdS/CFT duality, which seems quite surprising to us. Indeed, this would imply that the gravitational path integral completely decouples baby universe observables from the rest of the theory. In other words, baby universe nucleation would not be able to arise from any bulk process that can be described by an ordinary holographic correspondence. However, we expect that a very high quantum matter density, which, for example, can occur inside of a black hole, may be able to bring a baby universe into the theory. In particular, this is clearly not the case in the axion model by Giddings--Strominger \cite{Giddings:1987cg}, where topology changes in the bulk are caused by matter fields, or even in the Euclidean model from string theory by Dijkgraaf--Gopakumar--Ooguri--Vafa \cite{Dijkgraaf:2006ab}, which is formulated in the context of AdS/CFT. More recently, some bulk topological operations coming from conformal field theory dynamics have been described by Marolf in \cite{Marolf:2019zoo}. If similar phenomena were to happen in realistic theories under consideration, they clearly would invalidate the third assumption.

One could argue that in our case, the definition of baby universes is a bit different: it requires an asymptotically AdS boundary and amounts to a change of dictionary. However, it still seems quite unlikely to us that no traditional process described by a bulk to boundary duality could be correlated in any way with the number of baby universes in the theory. Such a statement would in particular lead to the conclusion that baby universe creation and annihilation would be a completely independent process that does not depend on any other type of physics. In our opinion, it is the third assumption that gives the most physical obstructions for the baby universe hypothesis to be true in realistic models of quantum gravity.

\section{Discussion}
\label{sec:discussion}

In this section, we summarize our approach to constructing baby universes utilizing the observables of quantum gravity to build the baby universe algebra and the associated Hilbert space. We further explain that the validity of the baby universe hypothesis lies in the Marolf--Maxfield construction, purity of the gravitational path integral, and the factorizability between boundary and baby universe observables. Based on our results, we further discuss on its implications in the context of AdS/CFT correspondence.

\subsection{Summary of results} \label{sec:summary}

In this paper, we showed that one could get a rigorous understanding of baby universe constructions by shifting the focus from the Hilbert space to the algebras of observables. In particular, in the context of holographic correspondences, the Marolf--Wall thought experiment led to the conclusion that the algebra of observables of quantum gravity in the bulk had to be enhanced by an extra tensor factor that does not have any boundary interpretation. The crucial point of our analysis is that this extra tensor factor is a {\it commutative} algebra. We note that strictly speaking, the commutative nature of an algebra of observables associated to connected sum and gluing operations is not limited to the holographic context, and should still hold more generally, for example, in cosmology. In particular, the Marolf--Wall thought experiment is not required to obtain this commutativity. 

The resulting superselected Hilbert space of baby universes can then be obtained from a GNS construction with respect to a particular reference state which corresponds to the gravitational path integral. In the case where the gravitational path integral is represented by a pure state (as a requirement in Theorem \ref{thm:BUpure}), the GNS representation is extremely collapsing, and yields a one-dimensional Hilbert space, which satisfies the baby universe hypothesis of McNamara--Vafa. In Theorem \ref{thm:BU}, we exhibit three physical conditions which, when combined, guarantee the validity of this hypothesis: commutativity of the observables, purity of the overall quantum gravity state, and absence of entanglement between the baby universe degrees of freedom and the other ones.

Pure states on an Abelian $C^\ast$-algebra correspond to characters, whose structures explain the ``miraculous cancellations" of the different bulk topologies, and the moduli space of such admissible bulk path integrals is isomorphic to the Gelfand dual of the baby universe algebra, which is a very small part of state space. For example, in the case of the topological model considered in Marolf--Maxfield, this Gelfand dual will be isomorphic to $\mathbb{R}$. When the baby universe algebra is generated by a group, the Gelfand dual is isomorphic to the unit circle, and a precise analogy can be drawn with the $\theta$-vacua of QCD.  As a result, the baby universe hypothesis (or purity requirement) can be understood as a swampland condition for the gravitational path integral, and singles out a very simple moduli space of baby universe theories.

There is a deep mathematical reason why the Abelian nature of the algebra of baby universe observables gives such powerful results, which traces back to the origins of the field of operator algebras. We witnessed that due to the properties of the Gelfand isomorphism, any commutative $C^\ast$-algebra can be described as an algebra of continuous functions on a topological space, while the space of states on this algebra corresponds to probability measures on this space. This is not a coincidence, as the theory of operator algebras was originally conceived as a generalization of probability and integration theory to the quantum, noncommutative case. Here, all baby universe observables commute with each other; hence, they form a classical system that can be described by the classical tools of continuous functions (the $C^\ast$-algebra) and probability measures (the Gelfand space of pure states). 

In ordinary calculus, both functions and measures are required in order to perform computations. Similarly, in the quantum case, it will be necessary to know both the generalized functions describing the system (the $C^\ast$-algebra of observables) and the generalized integration measures (states). The GNS representation then gives a way to quotient out the irrelevant parts of function space that the state is blind to: in the commutative case, these parts correspond to functions whose support on the Gelfand space are disjoint from the support of the probability measure. The resulting Hilbert space will then depend on the $C^\ast$-algebra, but also on the reference state: it is a \textit{state-dependent} object. This state dependence also appears in the von Neumann algebra of operators in this GNS Hilbert space that results from a weak operator completion of the original $C^\ast$-algebra.\footnote{For more on von Neumann algebras and state dependence, see \cite{MonicaElliott2}.} We expect that, in order to describe state-dependent situations in quantum gravity, both $C^\ast$-algebras and reference states will play a role, and will have to be interwoven through GNS representations and von Neumann completions. The baby universe construction is the simplest occurrence of such phenomena, as it shows the power of the GNS representation in a purely classical setup.

The idea of working at the level of the algebras of observables, rather than the Hilbert spaces, seems to be the key to resolve the the baby universe hypothesis problem: adding the possibility of baby universes will enlarge the bulk algebra of observables and apparently break the traditional AdS/CFT duality; however, if the gravitational path integral is chosen in such a way that the baby universe hypothesis holds, then this subtle difference vanishes at the level of Hilbert spaces, as the extra tensor factor becomes trivial, and AdS/CFT is restored. To us, this is one of the many examples demonstrating that to formulate a complete theory of quantum gravity, one should not work directly at the level of the Hilbert spaces, but rather at the level of the $C^\ast$-algebra of observables. For example, in order to formulate entanglement wedge reconstruction in a consistent way for systems with operator pushing such as the infinite-dimensional HaPPY code, in contrast to state-pushing in simpler tensor network models \cite{Kang:2019dfi,Kang:2018xqy}, we have shown that a very similar GNS technique has to be employed \cite{MonicaElliott2,MonicaElliott}. Similarly, exact and approximate theorems for entanglement wedge reconstruction can be made much more natural by reformulating the problem directly in terms of algebras of observables, as argued in \cite{MonicaElliott:approxthm}. 

Another lesson to learn is that the GNS representation can sometimes very powerfully collapse a Hilbert space. A particularly striking example is the Marolf--Maxfield topological model, in which the GNS construction collapses the continuous space $\mathbb{R}$ of $\alpha$-states to the discrete space $\mathbb{Z}$. The case of the baby universes was, in a way, the simplest to treat, because the relevant $C^\ast$-algebra is Abelian. However, it would also be an interesting problem to understand the irreducible representations of more general non-Abelian $C^\ast$-algebras, as well as the corresponding moduli spaces of pure states, in order to understand how collapsing the dimension of the Hilbert space (or equivalently, the GNS representation) can be in more general contexts, and how powerful swampland-like conditions can arise from a purity constraint in these cases. We expect the general theory of quantum groups to be helpful in this more general context.

\subsection{On the physical meaning of the Marolf--Maxfield construction and the baby universe hypothesis}

In this paper, we proved that in the context of AdS/CFT, the baby universe hypothesis could be inferred from three conditions: the validity of the Marolf--Wall thought experiment (which also underpins the Marolf--Maxfield construction), the purity of the gravitational path integral as a whole, and its factorization between boundary observables and baby universe observables. 

In section \ref{sec:assumptions}, we argued that it was far from obvious that these assumptions were satisfied in a holographic theory of quantum gravity. While the validity of the Marolf--Wall thought experiment can potentially be inferred from a hypothetical approximate result on entanglement wedge reconstruction near a black hole horizon, and it could turn out that swampland arguments impose realistic theories of quantum gravity to correspond to pure states in high enough dimension, the factorization property leads to conclusions that seem quite odd. In particular, it would imply that no interaction is ever allowed between the baby universe observables and the boundary observables.

This main purpose of this paper was to clarify the mathematical structures underlying the Marolf--Maxfield construction and the baby universe hypothesis. Therefore, we remain largely agnostic about their respective physical meanings, although at least the baby universe hypothesis seems quite at odds with intuition.

\subsection{Does AdS/CFT need to be modified?}

It is a controversial question whether the baby universe construction requires the traditional AdS/CFT correspondence to be modified in the presence of a varying number of boundaries.

At the level of the $C^\ast$-algebras of observables, the Marolf--Wall thought experiment teaches us that the formation of a baby universe, or equivalently, allowing a new boundary to form wormholes with other bulks, cannot come from a CFT observable, and that an extra tensor factor has to be added to the theory. In this sense, AdS/CFT does need to be modified.

However, what the baby universe hypothesis teaches us is that this modification might disappear at the level of the Hilbert space. Indeed, if the GNS representation of the baby universe algebra is of dimension one, i.e. if the restriction of the gravitational path integral to the baby universe algebra corresponds to a pure state, the extra tensor factor becomes trivial once the GNS representation is performed. However, if the baby universe hypothesis turns out to not be true, the GNS representation of the baby unvierse Hilbert space will not have dimension one, and the AdS/CFT duality needs to be modified even at the level of Hilbert spaces, in the presence of a varying number of boundaries.

\section*{Acknowledgments}
The authors are grateful to Juan Felipe Ariza, Vincent Chen, Craig Lawrie, Matilde Marcolli, and Hirosi Ooguri for discussions. M.J.K. is thankful to Caltech arXiv meeting hosted by Hirosi Ooguri for generating interest and discussions. M.J.K. is supported by a Sherman Fairchild Postdoctoral Fellowship. This material is based upon work supported by the U.S. Department of Energy, Office of Science, Office of High Energy Physics, under Award Number DE-SC0011632. 
E.G. is funded by ENS Paris and would like to thank Matilde Marcolli for her guidance and constant support.
Research at Perimeter Institute is supported in part by the Government of Canada through the Department of Innovation, Science and Economic Development Canada and by the Province of Ontario through the Ministry of Colleges and Universities.

\end{document}